\documentclass[aps,prd,twocolumn,superscriptaddress,amsmath,showpacs,floatfix,nofootinbib,preprintnumbers]{revtex4-1}

\usepackage{graphicx}
\usepackage{dcolumn}
\usepackage{bm}
\usepackage{natbib}
\usepackage{multirow}
\usepackage[latin1]{inputenc}
\usepackage{hyperref}

\newcommand{\bad}{\begin{array}{ccc}}
\DeclareMathAlphabet{\mathsc}{OT1}{cmr}{m}{sc}

\usepackage{color}

\def\be{\begin{equation}}        
\def\ee{\end{equation}} 
\def\bear{\be\begin{array}}       
\def\eear{\end{array}\ee} 
\def\bea{\begin{eqnarray}} 
\def\eea{\end{eqnarray}}

\def\beqa{\begin{eqnarray}}
\def\eeqa{\end{eqnarray}}

\def\beq{\begin{equation}}
\def\eeq{\end{equation}}
\def\ba{\begin{array}}
\def\ea{\end{array}}

\def\bold#1{\setbox0=\hbox{$#1$} 
     \kern-.025em\copy0\kern-\wd0 
     \kern.05em\copy0\kern-\wd0 
     \kern-.025em\raise.0433em\box0 }

\def\vev#1{\left\langle #1\right\rangle}

\def\lnv{lepton number violation }

\newcommand {\ignore}[1]{}

\def\e6{$\mathrm{E(6)}$ }
\def\10{$\mathrm{SO(10)}$ }
\def\21{$\mathrm{SU(2)_L \otimes U(1)_Y}$ }
\def\31{$\mathrm{SU(3)_c \otimes U(1)_Q}$ }
\def\SM{$\mathrm{SU(3)_c \otimes SU(2)_L \otimes U(1)_Y}$ }

\def\3211{$\mathrm{SU(3) \otimes SU(2)_L \otimes U(1)_R \otimes U(1)_{B-L}}$ }
\def\321{$\mathrm{SU(3) \otimes SU(2) \otimes U(1)}$ }
\def\422{$\mathrm{SU(4) \otimes SU(2) \otimes SU(2)_R}$ }

\def\lsim{\mathrel{\rlap{\lower4pt\hbox{\hskip1pt$\sim$}}
    \raise1pt\hbox{$<$}}}         
\def\gsim{\mathrel{\rlap{\lower4pt\hbox{\hskip1pt$\sim$}}
    \raise1pt\hbox{$>$}}}         

\makeatletter
\renewcommand{\fnum@table}{\textbf{\tablename~\thetable}}
\renewcommand{\fnum@figure}{\textbf{\figurename~\thefigure}}
\makeatother

\def\mxth{\mathsurround=0pt }
\def\xversim#1#2{\lower2.pt\vbox{\baselineskip0pt \lineskip-.5pt
  \ialign{$\mxth#1\hfil##\hfil$\crcr#2\crcr\sim\crcr}}}             
\def\gtrsim{\mathrel{\mathpalette\xversim >}}                                    
    
\def\be{\begin{equation}}
\def\ee{\end{equation}}
\def\bea{\begin{eqnarray}}
\def\eea{\end{eqnarray}}

\newcommand{\AHEP}{AHEP Group, Instituto de F\'{\i}sica Corpuscular --
  C.S.I.C./Universitat de Val{\`e}ncia \\
  Campus de Paterna, Apt 22085, E--46071 Val{\`e}ncia, Spain}

\newcommand{\aap}{A\&A}
\newcommand{\apjl}{ApJL }

\newcommand{\jcap}{JCAP }
\newcommand{\mnras}{MNRAS }
\newcommand{\pasa}{PASA }

\definecolor{darkgreen}{RGB}{0,170,0}

\usepackage[normalem]{ulem}

\usepackage[usenames]{xcolor}

\begin{document}

\title{Updated CMB,  X- and $\gamma$--ray constraints on majoron dark matter}

\author {Massimiliano Lattanzi}
\email{lattanzi@fe.infn.it}
\affiliation{Dipartimento di Fisica e
  Science della Terra, Universit\`a di Ferrara and INFN, sezione di
  Ferrara, Polo Scientifico e Tecnologico - Edificio C Via Saragat, 1,
  I-44122 Ferrara Italy}

\author{Signe Riemer-S{\o}rensen} \affiliation{School of Mathematics
  and Physics, University of Queensland, St Lucia, Brisbane 4072,
  Queensland, Australia }

\author{Mariam T\'ortola}
\affiliation{\AHEP} 

\author{J. W. F. Valle}
\homepage{http://astroparticles.ific.uv.es/}
\affiliation{\AHEP}

\begin{abstract}
  The majoron provides an attractive dark matter candidate, directly
  associated to the mechanism responsible for spontaneous neutrino
  mass generation within the standard model \SM framework. Here we
  update the cosmological and astrophysical constraints on majoron
  dark matter coming from Cosmic Microwave Background (CMB)
  and a variety of X- and $\gamma$--ray observations.
\end{abstract}
 

\preprint{IFIC/13-15}

\maketitle

\section{Introduction \label {sec:intro}}

It is by now established that only a small fraction (less than 20\%)
of the matter in the Universe is in the form of ordinary - i.e.,
baryonic - matter, while the rest is in the form of so-called ``dark
matter".  The existence of dark matter is inferred by gravitational
anomalies at very different scales, ranging from galactic scales to
cluster scales, and all the way up to the cosmological scales.  In
particular, the 9-year data from the Wilkinson Microwave Anisotropy
Probe (WMAP) \cite{Hinshaw:2012fq,Bennett:2012fp} have provided even
stronger support to the six-parameter $\Lambda$CDM model, although
small-scale experiments like the Atacama Cosmology Telescope (ACT)
\cite{Sievers:2013wk} and the South Pole Telescope (SPT)
\cite{Hou:2012xq} hint to interesting, albeit discordant, deviations
from this simple picture, like e.g. the presence of other relativistic
degrees of freedom, in addition to the standard model neutrinos, or
deviations from ordinary gravity
\cite{Riemer-Sorensen:2013,Riemer-Sorensen:2013r}.  On the other hand,
the newly published results from the Planck satellite have provided an
even stronger support to the minimal $\Lambda$CDM model
\cite{Ade:2013uq}.

If the $\Lambda$CDM model is certainly a phenomenological success,
nevertheless it is puzzling from the theoretical point of view in many
aspects. On one side, the nature of both dark matter and dark energy,
that together make up for more than 95\% of the total energy budget of
the Universe, is still unknown. On the other side, the theory of
inflation, that explains the formation of the primeval seeds for
density fluctuations from which galaxies originate, is still waiting
to be embedded in a more fundamental theory. All these puzzles hint,
and possibly have their solution in, some physics beyond the standard
model (SM) of particle physics, or maybe in some modification of
general relativity.

Although its precise nature is still unknown, there is no shortage of
candidates for the role of dark matter. One of the most widely studied
candidates to date is the supersymmetric neutralino. Recent results
from the Large Hadron Collider (LHC), however, have greatly reduced
the available parameter space for supersymmetry, at least in its
simplest minimal supergravity implementations~\cite{Caron:2011ge},
reducing, perhaps, the appeal of supersymmetric dark matter
candidates. Other possible candidates include axions, Kaluza-Klein dark matter, keV dark matter,
such as sterile neutrinos, and many others. In particular, 
dark matter in the keV range has been advocated by many authors (see for example
Refs.~\cite{deVega:2012jm,Boyarsky:2009,Drewes:2013} and references
therein) as a possible solution for the shortcomings of the cold dark matter scenario
at small scales.

The evidence for the existence of dark matter is very strong, but
only limited to effects related to its gravitational interaction. The
search for non-gravitational evidence of the dark matter continues, in
the form of direct and indirect detection experiments, and by looking
for dark matter production in accelerators like the LHC.  A precise
underpinning of the dark matter and determination of its properties
can only come through a combination of these approaches.
  
If the dark matter has any connection to the world of SM particles,
there will be astrophysical signals one can search for, in particular
high energy photons from annihilating or decaying dark matter (see
Refs. \cite{Bringmann:2012a,Boyarsky:2012r} for a recent review).  The
most studied scenarios are the broad spectrum annihilation signals
from neutralinos, but the real smoking gun is line emission (either
directly from the decay/annihilation \cite{Bergstrom:1997,Bern:1997}
or from internal bremsstrahlung \cite{Beacom:2005,Bringmann:2008}),
for which the spectral and spatial distribution is not easily mimicked
by astrophysical sources. The recent claim of line emission at
$E_\gamma=130\, \mathrm{GeV}$ in the Fermi data
\cite{Bringmann:2012,Weniger:2012} has spurred a renewed interest in
emission line searches at high energy. However, the origin of the
possible signal at $E_\gamma=130\, \mathrm{GeV}$ is still unknown and
caution is encouraged with respect to its interpretation
\cite{Boyarsky:2012}.

It has been long suggested that the origin of dark matter could be
related to the origin of neutrino masses
\cite{gelmini:1984pe,berezinsky:1993fm}. In fact, the smallness of
neutrino masses, as compared to the other SM particles, is puzzling in
itself.  Most likely it is associated to the properties of the
messenger states whose exchange is responsible for inducing them. This
is the idea underlying the so-called seesaw
mechanism~\cite{gell-mann:1980vs,yanagida:1979,mohapatra:1980vn,Schechter:1980gr,Lazarides:1980nt},
whose details remain fairly elusive. Especially appealing is the
possibility that neutrino masses arise from the spontaneous violation
of ungauged lepton number~\cite{chikashige:1981ui,Schechter:1981cv}.
The associated Nambu-Goldstone boson, the majoron, could acquire a
mass from non-perturbative gravitational
effects~\cite{coleman:1988tj,kallosh:1995hi}, and play the role of the
dark matter particle. In Ref.~\cite{Lattanzi:2007ux} the viability of
the majoron as a dark matter particle was explored using the WMAP
3-year data and in Ref.~\cite{bazzocchi:2008fh} the possible X-ray
signature associated to majoron decay was investigated. A specific
theoretical model implementing the seesaw mechanism and an $A_4$
flavour symmetry was described in \cite{Esteves:2010sh}.

In this paper, we update our previous constraints in the light of the
more recent cosmological and astrophysical data. Regarding cosmology
we use the WMAP 9-year data \cite{Hinshaw:2012fq,Bennett:2012fp} (as
discussed in Sec. \ref{sec:CMB}, we do not expect our results to change significantly
using other CMB data).  On the astrophysical front we
include emission line searches on the entire range of photon energies
between $0.07\, \mathrm{keV}$ and $200\, \mathrm{GeV}$ from {\it
  Chandra} X-ray Observatory, X-ray Multi-Mirror Mission - Newton
({\it XMM}), High Energy Astronomy Observatory Program ({\it HEAO}),
INTErnational Gamma-Ray Astrophysics Laboratory ({\it INTEGRAL}),
Compton Gamma Ray Observatory ({\it CGRO}), and the Fermi Gamma-ray
Space Telescope.

The paper is organized as follows. In Sec. \ref{sec:majphys}, we
briefly recall the relevant majoron physics. In Secs. \ref{sec:CMB}
and \ref{sec:xgamma}, we derive observational constraints on the
majoron decay to neutrinos and photons, respectively, and we compare
them to the predictions of a general seesaw model. Finally, in
Sec. \ref{sec:conc} we draw our conclusions.

\section{Seesaw majoron physics \label{sec:majphys}}

The basic idea of majoron physics is that the lepton number symmetry
of the standard \SM model is promoted to a spontaneously broken
symmetry \cite{chikashige:1981ui,Schechter:1981cv}.
This requires the presence of a lepton-number-carrying complex scalar
singlet, $\sigma$, coupling to the singlet neutrinos $\nu^c_L$, as
follows,
\begin{equation}\label{eq:singlet}
 \lambda \sigma  {\nu^c_L}^T \sigma_2 \nu^c_L + H.c.
\end{equation}
with the Yukawa coupling $\lambda$. This term provides the large mass
term in the seesaw mass matrix
\begin{equation}
\label{ss-matrix} 
{\mathcal M_\nu} = \left[\begin{array}{cc}
    Y_3 v_3 & Y_\nu v_2 \\
    {Y_\nu}^{T} v_2  & Y_1 v_1 \\
\end{array}\right]
\end{equation}
in the basis of ``left'' and ``right''-handed
neutrinos  $\nu_{L}$, $\nu^{c}_{L}$. The model is characterized by singlet, doublet and triplet
Higgs scalars whose vacuum expectation values (vevs) are arranged to
satisfy $v_1 \gg v_2 \gg v_3$ obeying a simple vev seesaw relation of
the type
 \begin{equation}
   v_3 v_1 \sim {v_2}^2.
 \label{eq:123-vev-seesaw}
 \end{equation}
 The vev $v_1$ drives \lnv and induces also a small but nonzero $v_3$,
 while ${v_2}$ is fixed by the masses of the weak gauge bosons, the
 $W$ and the $Z$. Note that the vev seesaw condition implies that the
 triplet vev $v_3 \to 0$ as the singlet vev $v_1 \to \infty$.  The
 three vevs determine all entries in the seesaw neutrino mass matrix.
 Regarding the Yukawa couplings, $Y_\nu$ is an arbitrary flavour
 matrix, while $Y_3$ and $Y_1$ are symmetric.  The effective light
 neutrino mass obtained by perturbative diagonalization of
 Eq.~(\ref{ss-matrix}) is of the form
\begin{equation}
  \label{eq:ss-formula0}
  m_{\nu} \simeq Y_3 v_3 -
Y_\nu {Y_1}^{-1} {Y_\nu}^T \frac{{{v_2}}^2}{v_1}
\end{equation}
Together with Eq.~(\ref{eq:123-vev-seesaw}) this summarizes the essence of the seesaw mechanism.


In order to identify which combination of Higgs fields gives the
majoron, J, one may write the scalar potential explicitly, minimize
it, and determine the resulting scalar mass matrices.
However, one can do this simply by exploiting the invariance
properties of the Higgs potential $V$~\cite{Schechter:1981cv}. 
The result is proportional to the combination
\begin{eqnarray}
  \label{eq:cp-J}
J &\propto&  v_3 {v_2}^2 \,\text{Im}(\Delta^0) - 2 v_2 {v_3}^2 \,\text{Im}(\Phi^0) \\
&& + v_1 ( {v_2}^2 + 4 {v_3}^2) \, \text{Im}(\sigma)  
\end{eqnarray}
up to a normalization factor. $\text{Im}()$ denotes the imaginary
parts, while $\Delta^0$ and $\Phi^0$ refer to the neutral components
of the triplet and doublet scalars respectively, and $\sigma$ is the
scalar singlet introduced in Eq. \ref{eq:singlet}.  We remark the
presence of the quartic lepton-number-conserving term
\begin{equation}
 \Phi^\dagger \Delta \tau_2 \Phi^*\sigma^* + H.c.   
\end{equation}
in the scalar potential. Here $\tau_2$ is the weak isospin Pauli
matrix, and $v_2\equiv\vev{\Phi}$, $v_1\equiv\vev{\sigma}$,
$v_3\equiv\vev{\Delta}$.  This term illustrates the need for mixing
among neutral fields belonging to all three Higgs multiplets in the
expression for the majoron, Eq.~(\ref{eq:cp-J}).
As a result the majoron has an explicit coupling to two photons
leading to a possible indirect detection of majoron dark matter by
searching for the corresponding high energy photons
\cite{bazzocchi:2008fh}, which we treat in Sec. \ref{sec:xgamma}.


We now turn to the form of the couplings of the majoron within the
above seesaw scheme, characterized by spontaneous \lnv in the presence of
singlet, doublet and triplet Higgs scalars.
Again one can derive the form of the couplings of the majoron using
only the symmetry properties, as described in
Ref.~\cite{Schechter:1981cv},
\begin{equation}
{\cal L}_{\rm Yuk}=
\frac{iJ}{2}{\displaystyle \sum_{ij}} \nu^T_{i} g_{ij} \sigma_2 \nu_{j} +H.c.
\end{equation}
The result is a perturbative expansion for the majoron couplings
\begin{equation}\label{eq:Jnunu} {g}_{ij}= -\frac{m_i^\nu}{v_1}
  \delta_{ij} + \dots
\end{equation}
where the dots $\dots$ denote higher order terms.
One sees that, to first approximation, the majoron couples to the
light mass-eigenstate neutrinos inversely proportional to the \lnv
scale $v_1\equiv \vev{\sigma}$ and proportionally to their mass.
With this we can compute the dark matter majoron decay rate to
neutrinos as
\begin{equation}
  \label{eq:42}
  \Gamma_{J\rightarrow\nu\nu} = \frac{m_J}{32\pi} \frac{\sum_i (m^\nu_i)^2}{2
    v_1^2} \ ,
\end{equation}
where the Majoron mass $m_J$ is presumably generated by
non-perturbative gravitational
effects~\cite{coleman:1988tj,kallosh:1995hi}.  Moreover, there is a
sub-leading majoron decay mode to photons. Within the general seesaw
model this decay is induced at the loop level, resulting in
\cite{bazzocchi:2008fh}
\begin{equation}
\label{eq:gg}
  \Gamma_{J\rightarrow\gamma\gamma} = \frac{\alpha^2 m_J^3}{64\pi^3}  \left| \sum_f N_f  Q_f^2
\frac{2 v_3^2}{v_2^2 v_1} (- 2 T_{3}^f)
    \frac{m_J^2}{12 m_f^2}\right|^2\,,
\end{equation}
where $N_f$, $Q_f$, $T_3^f$ and $m_f$ denote respectively the color
factor, electric charge, weak isospin and mass of the SM electrically
charged fermions $f$.  We note that this formula is an approximation
valid for $m_J \ll m_f$; however we will always use the exact formula
in the actual calculations.

The decay of the majoron dark matter to neutrinos provides the most
essential and model--independent feature of the majoron dark matter
scenario, namely, it is a decaying dark matter model where the majoron
decays mainly to neutrinos, a mode that is constrained from the CMB
observations, as we discuss in Sec. \ref{sec:CMB}. 

\section{CMB constraints on the invisible decay $J\to\nu\nu$ 
\label{sec:CMB}}

\subsection{Method}

We start by deriving constraints on the majoron properties from CMB
anisotropy data.  The majoron differs from most dark matter candidates
in that it is unstable, since it must decay to neutrinos, as seen in
Eq.~(\ref{eq:42}), although it obviously has to be long-lived enough
to play the role of the dark matter today.

In order to investigate the observable effects of majoron decay on the
CMB, the Boltzmann equation describing the phase-space evolution of
dark matter particles must be modified accordingly, as shown e.g.  in
Ref.  \cite{kaplinghat:1999xy}, both at the background and at the
perturbation level.  The main effect of the late dark matter decay to
invisible relativistic particles is an increase of the late
integrated Sachs-Wolfe effect, caused by the presence of an extra
radiation component at small redshifts. This is reflected in the CMB
power spectrum by an increased amount of power at the largest angular
scales (i.e., small multipoles). Too large a decay rate would produce
too much radiation and too much large-scale power, and thus be at
variance with observations.  In Ref.  \cite{Lattanzi:2007ux} two of us
have used this effect to constrain the majoron lifetime. However, we
did not take properly into account the effect of majoron decay on the
age of the Universe; this led to an underestimate of the upper limit
of the majoron decay rate. We have now corrected this.

We use a modified version of CAMB \cite{Lewis:1999bs}, taking into
account the finite lifetime of the majoron, to evolve the cosmological
perturbations and compute the anisotropy spectrum of the CMB for given
values of the cosmological parameters. We assume that we can neglect
the velocity dispersion of majorons, i.e. we treat the majoron as a
cold dark matter particle ($m_J \gg T_J$). In order to compute
bayesian confidence intervals and sample the posterior distributions for the
parameters of the model, given some data, we use the
Metropolis-Hastings algorithm as implemented in CosmoMC
\cite{Lewis:2002ah} (interfaced with our modified version of CAMB). The model 
can be completely characterized by the six parameters of the standard $\Lambda$CDM model, namely
the present density parameters
$\Omega_\mathrm{b} h^2$ and $\Omega_\mathrm{dm} h^2$ of baryons and
dark matter respectively, the angular size of the sound horizon at recombination\footnote{We
have also repeated our analysis using $H_0$ as a base parameter in place of $\theta$ and 
found excellent agreement between the results obtained using the two parameterizations.}
 $\theta$, the optical depth
to recombination $\tau_\mathrm{rec}$, the spectral index $n_s$ and
amplitude $A_s$ (evaluated at the pivot scale $k_0=0.002$ Mpc$^{-1}$)
of the spectrum of primordial scalar fluctuations, to which we add the decay rate
$\Gamma_{J\to\nu\nu}$ of majorons to neutrinos. We marginalize
over the amplitude of the contamination from the Sunyaev-Zel'dovich signal.
We assume spatial flatness, massless neutrinos and adiabatic initial conditions. 

Following our previous work \cite{Lattanzi:2007ux}, instead of $\Omega_\mathrm{dm} h^2$
we use, as a base parameter, the quantity $s_\mathrm{early}$ defined as:
\begin{equation}
s_\mathrm{early} \equiv \left. \frac{\rho_\mathrm{dm}}{\rho_\mathrm{b}}\right|_{t\ll \Gamma^{-1}_{J\to\nu\nu}} \, ,
\label{eq:searly}
\end{equation}
\emph{i.e.}, the ratio  between the energy densities of dark matter and baryons at early times. This can
be related to the present dark matter density by means of 
\begin{equation}
\Omega_\mathrm{dm} h^2 = s_\mathrm{early} \, \Omega_\mathrm{b} h^2 \, e^{- \Gamma^{-1}_{J\to\nu\nu} t_0} \, ,
\end{equation}
where $t_0$ is the present age of the Universe.
Also, we do not vary directly the decay rate $\Gamma_{J\to\nu\nu}$ in our Monte Carlo runs,
but instead the ratio $\Gamma_{J\to\nu\nu}/H_0$. Finally, we express the amplitude of primordial fluctuations in
terms of $\ln (10^{10} A_s)$. Our base parameter set, consisting of those parameters
with uniform priors that are varied in the Monte Carlo runs, is summarized in the upper part of 
Tab. \ref{tab:priors}.

\begin{table*}[htdp]
\caption{Cosmological parameter used in the analysis. The upper part of the table lists the base parameters, \emph{i.e.}, those with
uniform priors that are varied in the Monte Carlo run. The lower part lists derived parameters of interest.
For each parameter, we quote the initial prior range (for base parameters only) and the confidence limits, in the form of posterior mean $\pm$ 
68\% uncertainty, with the exception of those parameter for which we can only derive an upper limit. In this case we only report the 95\% confidence limit.}
\begin{center}
\begin{tabular}{l l c c} 
Parameter				&	\quad Definition			&Prior range & \qquad Limits \\ \hline \hline
$\Omega_\mathrm{b} h^2$ 	&\quad  Present density of baryons		& [0.005, 0.1] & \qquad $0.02290 \pm 0.00054$ \\
$s_\mathrm{early}$	& \quad Primordial dark matter to baryon ratio\footnote{See definition in Eq. (\ref{eq:searly}).}	\quad & [0, 10] & \qquad $4.92 \pm 0.27$\\
$100\theta$					& \quad $100 \,\times$ angular size of the sound horizon at recombination \quad		& [0.5, 10] & \qquad $1.0401 \pm 0.0023$\\
$\tau_\mathrm{rec}$			& \quad Optical depth to recombination	& [0.01, 0.8] & \qquad $0.090 \pm 0.014$ \\
$n_s$					& \quad Spectral index of scalar perturbations 			&  [0.5, 1.5] & \qquad $0.977 \pm 0.014$ \\
$\ln(10^{10} A_s)$					& \quad Log amplitude of scalar perturbations at $k_0=0.002$ Mpc$^{-1}$ \qquad & [2.7, 4.0] & \qquad$3.162 \pm 0.048$ \\
$\Gamma_{J\to\nu\nu} H_0^{-1}$		&\quad  Ratio between majoron decay rate and expansion rate \qquad & [0, 1] & \qquad $< 0.269 $ \\[0.1cm]
\hline
$\Omega_\mathrm{dm} h^2$ & \quad Present dark matter density & $\dots$ & \qquad $0.102 \pm 0.010$ \\
$\Omega_\Lambda$ & \quad Present dark energy density\footnote{We consider a constant equation of state $w = -1$.} & $\dots$ & \qquad $0.743 \pm 0.030$\\
$H_0$ & \quad  Hubble parameter today (km s$^{-1}$ Mpc$^{-1}$) & $\dots$ & \qquad $71.5 \pm 2.6$ \\
$\Gamma_{J\to\nu\nu}$ & \quad Majoron decay rate to neutrinos ($10^{-19}$s$^{-1}$) & $\dots$ & \qquad $< 6.40$\\
$m_J^{\mathrm{eff}}$ & \quad Effective majoron mass\footnote{See definition in the text.} (keV) & $\dots$ & \qquad $0.1577\pm 0.0067$ 
\end{tabular}
\end{center}
\label{tab:priors}
\end{table*}%

We perform our analysis using the most recent WMAP 9-year temperature
and polarization data \cite{Hinshaw:2012fq,Bennett:2012fp}. In
particular, for the temperature power spectrum we include data up to
$\ell_\mathrm{max}=1200$. We use the latest (V5) version of the WMAP
likelihood code, publicly available at the lambda
website\footnote{http://lambda.gsfc.nasa.gov/}.

\subsection{Results}

We first perform a control $\Lambda$CDM run by fixing the value of
$\Gamma_{J\to\nu\nu}$ to 0 (i.e., we consider stable dark matter) and
check that we can consistently reproduce the results quoted in the
WMAP9 parameter paper \cite{Hinshaw:2012fq}. Then we allow for the
possibility of decaying dark matter; our results for the cosmological parameters 
are summarized in the fourth column of Tab. \ref{tab:priors}.
We find that the limits on the
parameters of the standard $\Lambda$CDM model do not change
significantly, with the one exception of the present dark matter
density. In particular, taking as a reference the values quoted in
Table 3 of Ref. \cite{Hinshaw:2012fq}, the uncertainties of the other
parameters increase by less than 10\%, and the posterior means shift
by a fraction of a standard deviation at most.

\begin{figure}
\begin{center}
\includegraphics[width=0.95\linewidth,keepaspectratio]{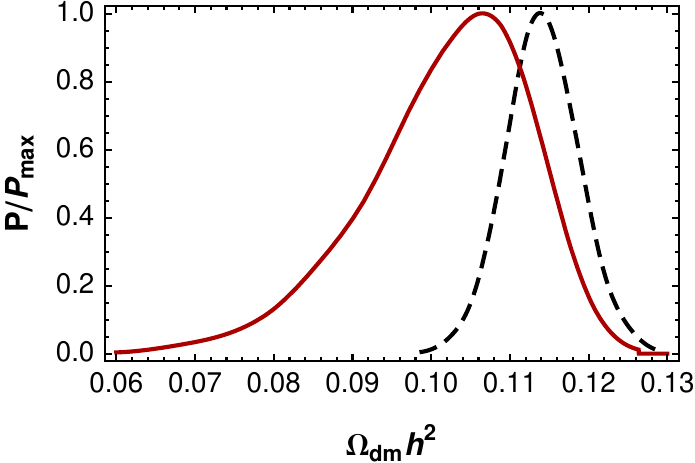}
\caption{One-dimensional posterior for the dark matter density
  parameter $\Omega_\mathrm{dm} h^2$ obtained from the WMAP9 data, for
  the $\Lambda$CDM (black dashed) and decaying majoron DM (red solid)
  models.}
\label{fig:omdmh2}
\end{center}
\end{figure}

For the present majoron density parameter, we find:
\begin{equation}
\Omega_\mathrm{dm} h^2 = 0.102 \pm 0.010 \qquad (68\%\,\mathrm{C.L.})\, .
\end{equation}
Compared with the WMAP9 $\Lambda$CDM result of $\Omega_\mathrm{dm} h^2
= 0.1138 \pm 0.0045$ \cite{Hinshaw:2012fq}, our estimate is shifted
towards smaller values, and has an uncertainty which is a factor two
larger. In Fig. \ref{fig:omdmh2}, we compare the marginalized
one-dimensional posterior for $\Omega_\mathrm{dm}h^2$ in the framework
of the decaying dark matter model, with the one obtained from our
control $\Lambda$CDM run.  The reason for both the shift and the
increase of the error bars will be discussed below.

For the purpose of the present analysis, we are mainly interested in the limits 
on the decay width of majoron to neutrinos, $\Gamma_{J\to\nu\nu}$. We get the 
following upper limit at 95\% C.L.
\begin{equation}
\Gamma_{J\to\nu\nu} \le 6.4\times 10^{-19}\,\mathrm{s}^{-1} \, .
\label{eq:gammanu}
\end{equation}
after marginalizing over the remaining parameters of the model. This
results in a lower limit to the majoron lifetime $\tau_J \ge
50\,\mathrm{Gyr}$, roughly four times the age of the Universe. This
limit is slightly relaxed to $\tau_J \ge 37\,\mathrm{Gyr}$ when we
allow for the possibility of extra degrees of freedom at the time of
recombination, by varying $N_\mathrm{eff}$.

\begin{figure*}
\begin{center}
\includegraphics[width=0.45\linewidth,keepaspectratio]{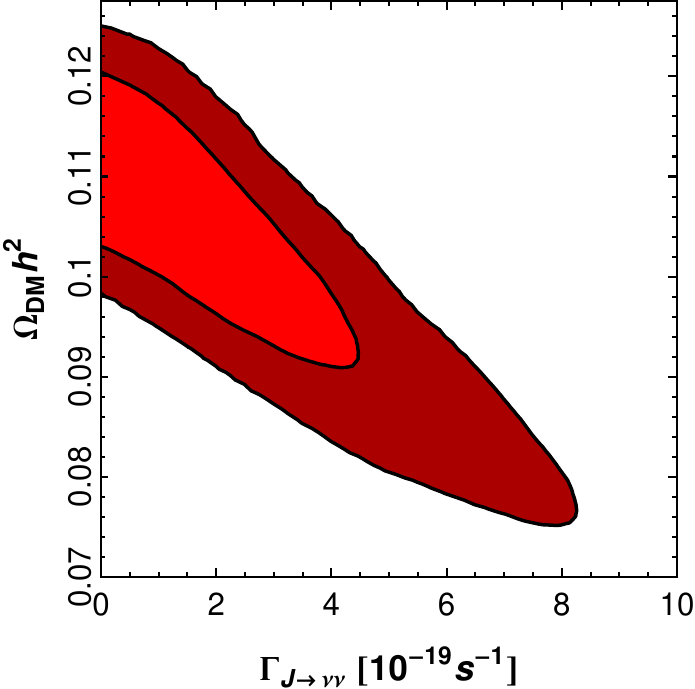}
\includegraphics[width=0.45\linewidth,keepaspectratio]{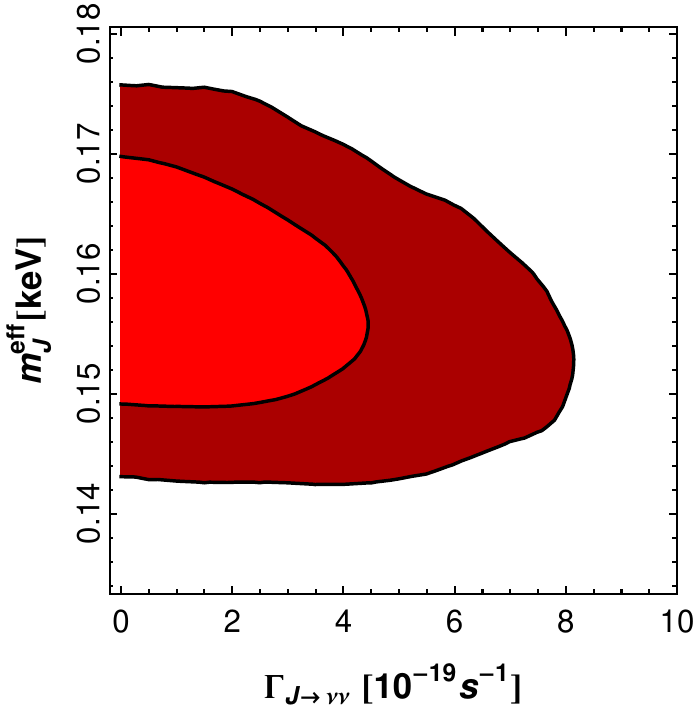}
\caption{Two-dimensional WMAP-9 constraints on the majoron dark matter
  parameters. The light (dark) shaded regions correspond to 68\%
  (95\%) confidence regions. Left panel: present density vs. decay
  rate to neutrinos. Right panel: effective mass vs. decay rate to
  neutrinos.}
\label{fig:2D_majoron}
\end{center}
\end{figure*}

In the left panel of Fig. \ref{fig:2D_majoron} we show 68\% and 95\%
confidence regions in the $(\Gamma_{J\to\nu\nu},\,\Omega_\mathrm{dm}
h^2)$ parameter plane. There is an evident anti-correlation between
decay rate and abundance that is explained by the fact, already
discussed in Ref. \cite{Lattanzi:2007ux}, that the CMB anisotropy
spectrum is mainly sensitive to the amount of dark matter prior to the
time of recombination (through the height of the first peak), as this
sets the time of matter-radiation equality.  Once the amount of dark
matter in the early Universe is fixed, increasing the decay rate
results in a smaller amount of dark matter at the present time, and
viceversa.  This degeneracy between $\Omega_\mathrm{dm}h^2$ and
$\Gamma_{J\to\nu\nu}$ explains the different shape of the posteriors
shown in Fig. \ref{fig:omdmh2}, and consequently explains the lower
value and larger uncertainty of the estimate of
$\Omega_\mathrm{dm}h^2$ with respect to those obtained for
$\Lambda$CDM. In fact, if we compute constraints on the {\it
  primordial} dark matter density (for example considering the
combination $\Omega_\mathrm{dm} h^2 \exp\left({\Gamma_{J\to\nu\nu}
    t_0}\right)$, which is, up to a multiplicative constant, the
comoving density of dark matter at early times), we find consistent
results between the $\Lambda$CDM and the majoron DM models.

In the limit of cold dark matter, one can not directly constrain the
mass of the dark matter particle itself, since this quantity never
appears explicitly neither in the background nor in the perturbation
equations.  Instead, the mass only appears implicitly inside the
physical density parameter $\Omega_\mathrm{dm} h^2$, in combination
with the present number density $n^0_\mathrm{dm}$, since for
nonrelativistic particles $\Omega_\mathrm{dm} h^2 \propto
\rho_\mathrm{dm} = m_\mathrm{dm} n^0_\mathrm{dm}$.  The calculation of
the number density relies on the knowledge of the production mechanism
of the dark matter particle and on its thermal history. If the majoron
was in thermal equilibrium with the rest of the cosmological plasma at
some early time, and decoupled while still relativistic, one finds
\begin{equation}
\Omega_\mathrm{dm}^\mathrm{th} h^2 = \left( \frac{g_{*S}}{106.75} \right)^{-1}\left(\frac{m_J}{1.40\,\mathrm{keV}}\right) e^{\Gamma_{J\to\nu\nu} t_0} \, ,
\label{eq:thermomega}
\end{equation}
where $g_{*S}$ parametrizes the entropy content of the Universe at the
time of majoron decoupling. If the majoron decouples when all the
degrees of freedom of the SM of particle physics are excited and in
thermal equilibrium, one has $g_{*S} = 106.75$.  In order to account
for a more general scenario, following Ref.  \cite{Lattanzi:2007ux},
we write
\begin{equation}
\Omega_\mathrm{dm}^\mathrm{th} h^2 = \beta \left(\frac{m_J}{1.40\,\mathrm{keV}}\right) e^{\Gamma_{J\to\nu\nu} t_0} \, ,
\label{eq:defbeta}
\end{equation}
so that $\beta =1$ corresponds to the case of a thermal majoron
decoupling when $g_{*S} =106.75$. The parameter $\beta$ encodes our
ignorance about the majoron thermal history, and $\beta \neq 1$ can
account both for a thermal majoron decoupling when $g_{*S} \neq
106.75$, or for a non-thermal distribution. Non-thermal
  production mechanisms include, for example, a phase transition
  \cite{berezinsky:1993fm} or the evaporation of majoron strings
  \cite{Rothstein:1992rh}. However, a detailed study relating the
  parameters of the underlying particle physics model to the
  cosmological majoron abundance in any of these scenarios is still
  lacking, so it is difficult to identify, on purely theoretical
  grounds, the range of reasonable values of beta.

Using Eqs. (\ref{eq:defbeta}), we
can constrain the ``effective mass'' $
m_J^{\mathrm{eff}}\equiv  \beta\, m_J$ and get:
\begin{equation}
m^\mathrm{eff}_J =  (0.158 \pm 0.007)\, \mathrm{keV}   \qquad (68\%\,\mathrm{C.L.})\, .
\end{equation}
In the right panel of Fig. \ref{fig:2D_majoron} we show 68\% and 95\%
confidence regions in the $(\Gamma_{J\to\nu\nu},\, m^\mathrm{eff}_J)$
plane. This should substitute the results appearing in
Ref.~\cite{Lattanzi:2007ux}. Moreover, we stress again that this
constraints can be read in terms of the actual majoron mass only in
the case of thermal majoron decoupling when $g_{*S} = 106.75$ (i.e.,
$\beta=1$). Since the CMB does not really constrain the majoron mass
(at least in the cold limit), in the next section we will consider
values of the mass also outside the keV range. We do not
  consider values of the mass below $\sim$ 0.15 eV (corresponding to
  $\beta\gtrsim 1$) as they are likely to lead to problems in the
  context of structure formation due to the large free-streaming
  length of the particle \cite{Boyarsky:2008xj} (although a detailed
  study would require the knowledge of the full distribution
  function). The soft X-ray band is also observationally challenging
  with no current high-resolution observations appropriate for
  line-searches.

\section{X- and $\gamma$-ray constraints on the photon decay $J\to
  \gamma\gamma$ \label{sec:xgamma}}

One of the most interesting features of spontaneous \lnv within the
general \SM seesaw model is that the neutrino decay mode in
Eq.~(\ref{eq:42}) is accompanied by a two-photon mode,
Eq.~(\ref{eq:gg}), as a result of the Eq.~(\ref{eq:cp-J}). The decay
into photons is constrained by a number of astrophysical observations.

\subsection{Existing constraints}

\begin{figure}
	\includegraphics[width=8.3cm]{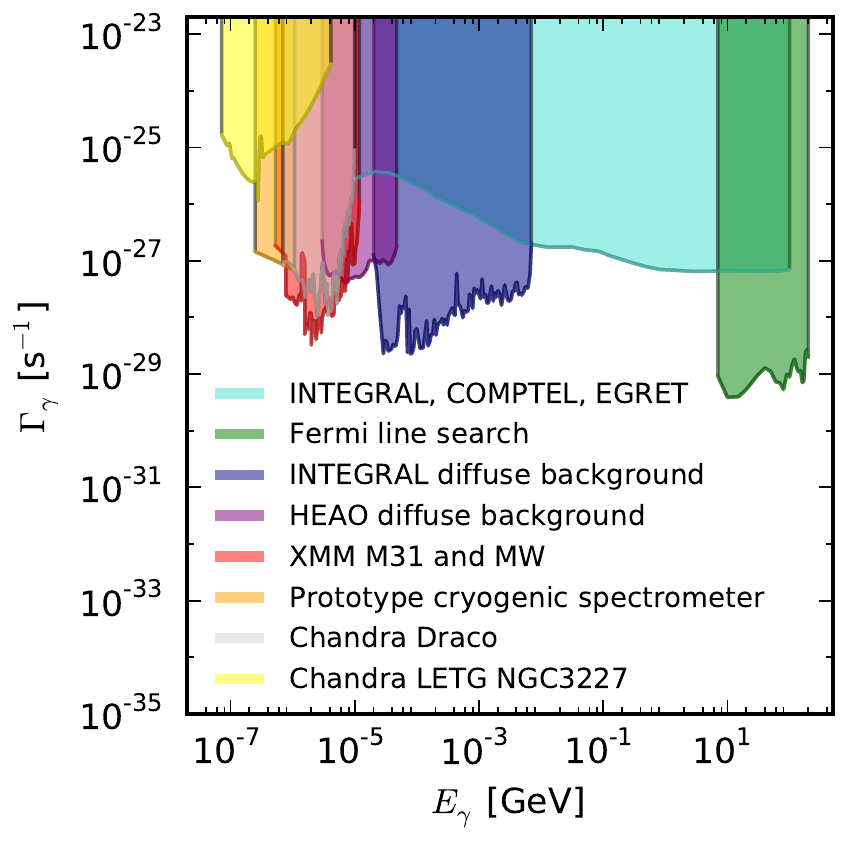} 
	\caption{$3\sigma$ line emission constraints on the decay rate
          into two mono-energetic photons. These constraints
            apply to all dark matter candidates with this
          signature. The constraints are taken from: yellow
          \cite{Bazzochi:2008}, orange \cite{Boyarsky:2007b}
          (conservatively rescaled by a factor of two due to mass
          estimate uncertainties as recommended in
          \cite{Boyarsky:2009}), red \cite{Boyarsky:2008b}, grey
          \cite{Riemer-Sorensen:2009}, purple \cite{Boyarsky:2006a},
          blue \cite{Boyarsky:2008a}, cyan \cite{Yuksel:2008b}, green
          \cite{Ackermann:2012}.}
	\label{fig:E_Gamma}
\end{figure}

In Fig. \ref{fig:E_Gamma} we plot the emission line constraints over
the wide range of photon energies of $0.07 \, \mathrm{keV}$ to $200 \,
\mathrm{GeV}$.

The very soft X-ray emission is covered by {\it Chandra} Low Energy
Transmission Grating (LETG) observations of NGC3227 ($0.07-4.1\,
\mathrm{keV}$) \cite{Bazzochi:2008} and a rocket borne light cryogenic
spectrometer ($0.25-1.1\, \mathrm{keV}$)
\cite{Boyarsky:2007b,McCammon:2002}.

The $0.3-12\, \mathrm{keV}$ range is well covered with constraints
from various objects observed with the {\it Chandra} and {\it XMM}
X-ray telescopes
\cite{Boyarsky:2006a,Boyarsky:2006b,Riemer-Sorensen:2006,Abazajian:2006a,Boyarsky:2007,Riemer-Sorensen:2007,Boyarsky:2008a,Boyarsky:2008b,Riemer-Sorensen:2009,Loewenstein:2009,Loewenstein:2010,Boyarsky:2010,Loewenstein:2012,Borriello:2012}.
In Fig. \ref{fig:E_Gamma} we have chosen the strongest robust
constraints\footnote{Some analyses have
  claimed stronger constraints in this energy interval, but were later found to be too
  optimistic. Ref. \cite{Abazajian:2001} underestimated the flux by
  two orders of magnitude
  \cite{Abazajian:2006a,Boyarsky:2006b}. According to
  Ref. \cite{Boyarsky:2008b} the mass was overestimated in
  Ref. \cite{Watson:2006} leading to too restrictive constraints. The
  constraints in Ref. \cite{Yuksel:2008} might be too restrictive due
  to the choice of source profile \cite{Boyarsky:2008a}, and the
  spectral resolution appears overestimated in
  Ref. \cite{Watson:2012}} from {\it XMM} observations of the Milky Way and M31
\cite{Boyarsky:2008b} and from {\it Chandra} observations of the Draco
dwarf galaxy \cite{Riemer-Sorensen:2009}.

The diffuse X-ray background observed with HEAO was searched for
line emission by Ref. \cite{Boyarsky:2006a} over the range $3-48\,
\mathrm{keV}$, and line emission constraints have been derived from
INTEGRAL SPI observations of the soft $\gamma$-ray background
($20\,\mathrm{keV} - 7\, \mathrm{MeV}$) \cite{Boyarsky:2008a}. For
energies above those covered by INTEGRAL the constraints are two
orders of magnitude worse as this range is only covered by a
combination of the rather old COMPTEL and EGRET instruments onboard
the CGRO. However, line emission
constraints have been derived up to $100\, \mathrm{GeV}$
\cite{Yuksel:2008b}. The most recent flagship for $\gamma$-ray
searches is the Fermi $\gamma$-ray Space Telescope, for which line emission
searches have been performed for the range of $7-200\, \mathrm{GeV}$
\cite{Ackermann:2012}.

\subsection{Future improvements}

The constraints on the majoron decay rate into two mono-energetic
photons shown in Fig. \ref{fig:E_Gamma} can be improved by increasing
the statistics or the spectral resolution. Increasing statistics
(either by exposure time or by sensitivity) improves the constraints
as $\Gamma^\mathrm{new}_{\gamma\gamma} =
\sqrt{N_{\gamma\gamma}^\mathrm{ex}/N_{\gamma\gamma}^\mathrm{new}}
\Gamma^\mathrm{ex}_{\gamma\gamma}$, where
$N_{\gamma\gamma}^\mathrm{ex,\, new}$ are the existing and new total
of photons per bin (assuming both source and background counts
increase by the same amount). Increasing the spectral resolution
improves the constraints directly as
$\Gamma_{\gamma\gamma}^\mathrm{new} =
E_\mathrm{FWHM}^\mathrm{new}/E_\mathrm{FWHM}^\mathrm{ex}\Gamma_{\gamma\gamma}^\mathrm{ex}$
and is consequently preferable but also technically more challenging.

\subsection{Model comparison}

We now compare the observational constraints obtained in the previous
section to the predictions of different realizations of a general
majoron seesaw model.  In particular, we perform a random scan over
the Yukawa matrices ($Y_\nu$, $Y_1$, $Y_3$) and vevs ($v_1$, $v_3$)
that characterize the seesaw mass matrix $\cal{M}_\nu$ in
Eq. (\ref{ss-matrix}). For each point in the parameter space we
evaluate the effective light neutrino mass matrix and the Majoron
decay rate to neutrinos following Eqs. (\ref{eq:ss-formula0}) and
(\ref{eq:42}), respectively.  We then choose, among all possible
realizations, those that are in agreement with current neutrino
oscillation data~\cite{Tortola:2012te} as well as with the bound on
neutrino decay rate in Eq. (\ref{eq:gammanu}). Finally we compute the
corresponding decay rate to photons, as described in
Sec. \ref{sec:majphys}.

\begin{figure}
	\includegraphics[width=8.3cm]{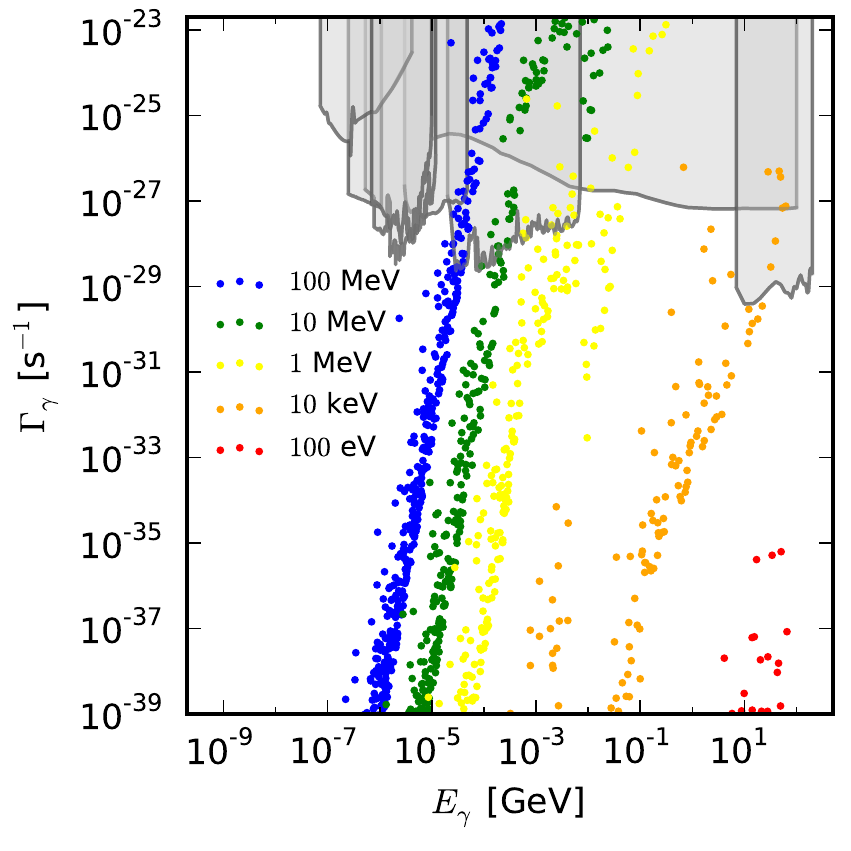} 
	\caption{The line emission constraints from Figure \ref{fig:E_Gamma} (grey) compared to model predictions (colored dots), for different values
	of the triplet vev $v_3$.}  
	\label{fig:models}
\end{figure}

We show the results of our scan in parameter space in
Fig. \ref{fig:models}, together with the constraints already shown in
Fig. \ref{fig:E_Gamma}.  It is clearly visible that the
$J\rightarrow\gamma\gamma$ constraints from line emission searches
already begin to cut the remaining parameter space for realistic
models.  This happens in particular for models with $v_3$ larger than
a few MeVs. However, models with lower values of the triplet vev
predict a photon flux that falls below the observational limits, as
seen from the figure. For example, for $v_3 < 100$~eV, predictions lie
below both current and planned $\gamma$-ray observatory sensitivities.

Note that, although for masses above 1 MeV the majoron could decay to
electron-positron pairs, nevertheless the branching ratio is
negligible. However, at even higher masses, new decay channels open up,
with the production of muon-anti-muon pairs etc. In this case these
decays would produce continuum gamma-ray emission at energies below
the emission line. This does not change any of the constraints given
in Figs. \ref{fig:E_Gamma} and \ref{fig:models}, though it would give
rise to additional constraints from continuum photon fluxes and
subsequent radio emission.

\section{Conclusions \label{sec:conc}}

We have updated previous constraints on the parameters of the majoron
dark matter model using the most recent CMB, X- and $\gamma$--ray
observations. From the CMB, we have derived an upper limit on the rate
of the invisible decay of the dark matter particle, namely, in the
framework of the model under consideration, on the majoron decay to
neutrinos. Translated in terms of the particle lifetime, this
constrains the majoron lifetime to be larger than 50 Gyrs.

Since, as already shown in Ref.~\cite{Lattanzi:2007ux}, the late decay
of dark matter mostly affects the large angular scale part of the CMB
power spectrum, where the uncertainty is dominated by cosmic variance,
we do not expect a dramatic improvement by using the Planck
data rather than WMAP9. Likewise, the small-scale data from ACT and
SPT are not expected to change significantly our constraints. However, we cannot exclude 
that a more precise determination of the intermediate to high-ell part of the spectrum could
affect, via parameter degeneracies, the estimation of the decay rate. We defer a more
careful study of this issue to a future work.

The majoron also possesses a subleading decay mode to two photons,
that can be constrained by astrophysical observations in the X and
$\gamma$ regions. We have compared these limits to the theoretical
predictions corresponding to different values for the parameters of
the underlying particle physics model.  We have found that the
observational constraints already exclude part of the parameter space
for models in which the vev of the triplet $v_3$ is larger than a few
MeVs. On the other hand, for smaller values of $v_3$, the current
limits need to be improved by at least 6 orders of magnitude before
the allowed region in parameter space can be reduced.

\section{Acknowledgments}
Work supported by MINECO grants FPA2011-22975 and MULTIDARK Consolider
CSD2009-00064, by Prometeo/2009/091 (Gen.  Valenciana), by EU ITN
UNILHC PITN-GA-2009-237920. The work of M.L. has been supported by
Ministero dell'Istruzione, dell'Universit\`a e della Ricerca (MIUR)
through the PRIN grant ``Galactic and extragalactic polarized
microwave emission'' (contract number PRIN 2009XZ54H2-002). The work
of M.T. is supported by CSIC under the JAE-Doc programme, co-funded by
the European Social Fund.

\bibliographystyle{apsrev4-1.bst}
\bibliography{majoron,additions_v2,merged,cosmo,newrefs-JR}
\end{document}